# Elimination of Interionic Hydrogen Bonding in the Imidazolium-Based Ionic Liquids

Vitaly V. Chaban[1]

(1) P.E.S., Vasilievsky Island, Saint Petersburg, Russian Federation.

**Abstract**. Hydrogen bonding is a phenomenon of paramount importance in room-temperature ionic liquids. The presence or absence of the hydrogen bond drastically alternates self-diffusion, shear viscosity, phase transition points, and other key properties of a pure substance. For certain applications, the presence of cation-anion hydrogen bonding is undesirable. In the present paper, we investigate perspectives of removing the hydrogen...fluorine interionic attraction in the imidazolium borates, the strongest non-covalent interaction in this type of system. Chemical modification of the tetrafluoroborate anion not only eliminates hydrogen bonding but also changes the most thermodynamically preferable orientation of the cation in the vicinity of the anion. Although the most acidic hydrogen atom of the imidazole ring remains the paramount electrophilic center of the cation, it does not engender a strong electrostatically driven coordination pattern with the properly modified anions. The reported new physical insights help compose more robust ionic liquids and tune solvation properties of the imidazolium-based RTILs.

**Graphical Abstract**

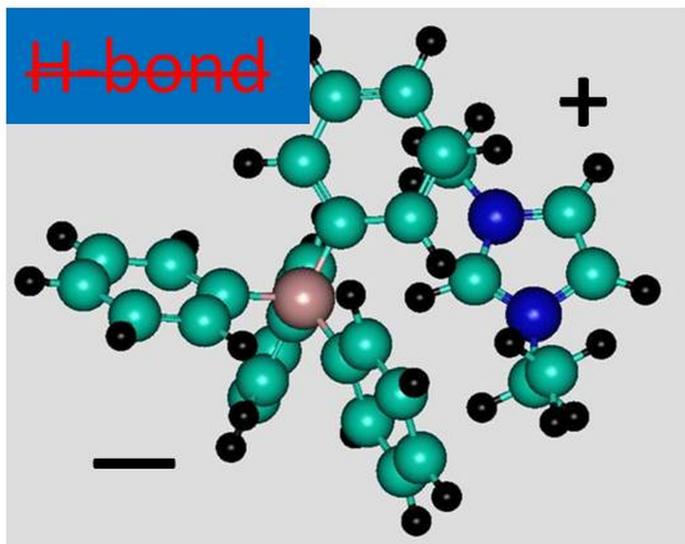

**Introduction**

Room-temperature ionic liquids (RTILs) are a general term to designate an extensive group of organic and organic-inorganic salts that exhibit melting points above 373 K.[1–3] RTILs possess somewhat specific physical-chemical properties compared to molecular liquids.[4] They conduct electricity, exhibit small vapor pressures, and demonstrate a robust catalytic activity. Furthermore, some RTILs are environmentally friendly and serve as rather universal solvents for inorganic and biological entities thanks to their amphiphilic structures and a wide liquid range.[5] Asymmetric structures of the cations and the anions coupled with the strongly delocalized excess and deficiency of the electron density are very important to deteriorate crystallization at higher temperatures. The tunability of the physical properties of RTILs by combining and functionalizing different cations and anions greatly enhances researchers' interest in this area of chemistry.

The well-recognized shortcomings of all RTILs in the context of industry-level applications are high shear viscosities and low self-diffusivities. These drawbacks come not only from more or less long hydrocarbon chains confining the central polarized atom but also from the relatively strong cation-anion coordination in the condensed phase. Nevertheless, the electrostatic attraction in RTILs is significantly weaker than in the case of inorganic molten salts.[6] The cation-anion hydrogen bonding constitutes another powerful factor that increases melting temperatures and shear viscosities of RTILs.[7] Hydrogen bonding can be avoided by employing weakly coordinating ions and chemically modifying the electron density distribution on the cation or the anion by introducing inert, non-polar, and bulky substituents.

The imidazolium-based ionic liquids[8–11] hold promise for application in the field of energy storage. For instance, they can be used as robust electrolytes in electric double-layer capacitors because the imidazolium-based RTILs possess a much wider potential window, better thermal stabilities, higher power densities, and lower saturated vapor pressures as compared to most of the conventional organic and aqueous electrolyte systems.[12] In the meantime, the imidazolium-based RTILs are inferior to their traditional competitors according to ionic conductivity. Ionic conductivity directly depends on the self-diffusivities of the ions, therefore both enumerated physical properties must be tuned together.

A common practical solution to the high viscosity problem is the preparation of mixtures containing ionic and molecular liquids in certain proportions.[4,13–16] Thanks to polar moieties in the structure of many polar solvents, RTILs normally exhibit a strongly negative solvation enthalpy while forming true solutions. The resulting systems exhibit intermediate physical-chemical properties being actually stable and fine-tuned hybrids of their parental systems. The dependence of ionic conductivity versus the molar fraction of RTIL contains a maximum that features interplay between the diffusivity of the ions and their quantity per a unit volume. The electrolytes composed of ionic and molecular liquids hold promise for versatile applications in electrochemical devices.[17]

The progress in organic chemistry led to the synthesis of the unusual cations and anions that are referred to as weakly coordinating ions.[18,19] Such ions do not contain any prospective coordination sites for the counterions, whereas any electron-deficient and electron-rich atoms are sterically hidden. Tetrabutylammonium,[20] tetrabutylphosphonium,[21] tetraoctylammonium,[19] tetrakis(pentafluorophenyl)borate,[22] tetrakis(4-fluorophenyl)borate[23] are a few noteworthy examples whose structures were reliably characterized and successful synthetic procedures were proven. Weakly coordinating ions give rise to organic salts with low melting temperatures and unprecedented conformational flexibilities. The success achieved with the elimination of the strong cation-anion coordination patterns motivates researchers to develop systematic pathways to further increase the liquid-state range of RTILs.

The present work is devoted to the elimination of the cation-anion hydrogen bonding due to stepwise chemical modification of the anion's structure. We consider four imidazolium ionic liquids: 1-methyl-3-ethylimidazolium tetrafluoroborate [EMIM][BF$_4$], 1-methyl-3-ethylimidazolium tetramethylborate [EMIM][Me$_4$B], 1-methyl-3-ethylimidazolium tetraethyl borate [EMIM][Et$_4$B], 1-methyl-3-ethylimidazolium tetrafluorophenylborate [EMIM][Ph$_4$B]. We identify peculiarities of structure and electron density distributions to characterize an effect of the anions with a more and less confined electrostatic charge (electron-rich interaction center). The obtained trend may serve as a guideline for selecting more suitable anions for novel and better performing RTILs.

**Methodology**

The systems corresponding to [EMIM][BF$_4$], [EMIM][Me$_4$B], [EMIM][Et$_4$B], and [EMIM][Ph$_4$B] were represented by a single ion pair. The [EMIM][Me$_4$B] ionic liquid was also simulated as two ion pairs to understand the role of a larger system in the interionic coordination.

The global minimum, local minima and transition points were found on the potential energy surface of the simulated systems according to the following procedure. First, the system was equilibrated at 300 K using PM7-MD simulations.[24] The equations of motion were integrated with a time-step of $5 \times 10^{-4}$ ps according to the algorithm of Verlet. At equal time intervals, the external kinetic energy equivalent to the immediate temperature of 1000 K was provided to the system by randomly increasing the linear velocities of the atoms to maintain Maxwell-Boltzmann distribution. The system was left to relax during 100 time-steps before the Cartesian coordinates of the configuration were saved. The excess energy was subsequently removed from the system by the Andersen thermostat,[25] the collision frequency parameter being set to 0.05.

The geometries of the recorded configurations (one hundred snapshots per chemical composition) were optimized using the steepest descent algorithm at the PM7[26–28] and B3LYP/6-31G(d)[29,30] levels of theory until a convergence criterion, $1\times10^{-3}$ kJ mol$^{-1}$ nm$^{-1}$, was satisfied. At the end of each geometry optimization, the vibrational frequencies were calculated. The absence of the imaginary frequency in the vibrational profile was interpreted as a minimum point on the potential energy surface. On the contrary, a single imaginary frequency indicated a transition point ionic configuration. A few imaginary frequencies indicated a higher-order saddle point. Global minimum is the lowest energy local minimum.

The electronic wave function of the systems was obtained via the Becke-3-Lee-Yang-Parr (B3LYP) hybrid density functional theory. The basis atomic functions were taken from the polarized atom-centered split-valence double-zeta basis set 6-31G(d). The molecular wave function convergence threshold was $10^{-4}$ kJ mol$^{-1}$.

Electronic structure calculations, geometry optimizations, molecular dynamics simulations were performed in GAMESS-2014,[31] MOPAC-2016,[12,13] PM7-MD,[33] and GMSEARCH[24] programs. The in-home tools use selected functions, constants and definitions from the ASE and SciPy.org computational chemistry libraries.[34–36] Manipulations with chemical entities, adjustment of the input and output formats, and preparation of artwork were done in VMD-1.9.1,[37] Gabedit-2.5,[38] and Avogadro-1.2.0.[39]

**Results and Discussion**

We start from the investigation of the potential energy surfaces (Figure 1) and the effect of the anion on the number of stationary points. In the [EMIM][BF$_4$] system, only two stationary points were identified, the less thermodynamically favorable one being +4.5 kJ mol$^{-1}$ inferior to the global minimum state. The global minimum ionic configuration is depicted in Figure 2. The major structural pattern in this system is a formation of the strong hydrogen...fluorine H-bond. Its energy capacity is confirmed by the bond length, 0.191 nm. This result is in perfect agreement with molecular dynamics and quantum chemistry reports that were published during the last two decades.[40–42] In turn, the [EMIM][Me$_4$B] system is more conformationally flexible (Figure 1). Note that a greater number of atoms and chemical bonds in a system generally enhance diversity of its potential energy surface. Furthermore, the absence of the cation-anion coordination pattern fosters the emergence of new stationary states.

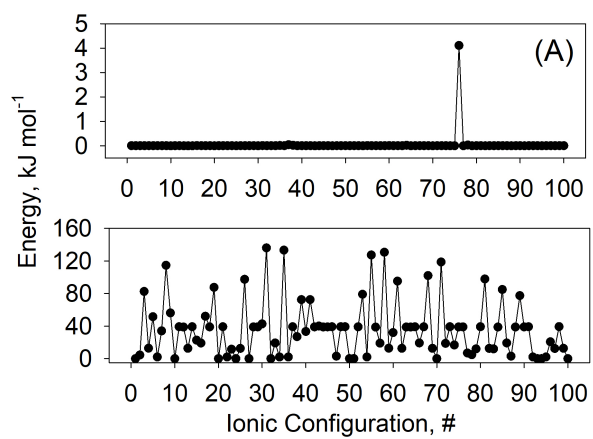

Figure 1. The potential energies of the stationary points found on the potential energy surface of (A) 1-ethyl-3-methylimidazolium tetrafluoroborate and (B) 1-ethyl-3-methylimidazolium tetramethylborate ion pairs. The energy of the global potential minimum is set to zero.

The stationary points with the potential energies of more than +50 kJ mol$^{-1}$ are transition states as suggested by the presence of significantly large imaginary vibrational frequencies. They correspond to the non-covalent rearrangements of the electrostatically interacting moieties of the cation and the anion and range between 100 and 400 cm$^{-1}$ in the complex space. In the present simulations, we did not aim to find transition states that would be responsible for chemical reactions, e.g. bond breakage (stretching). A more aggressive perturbation must be used to locate the mentioned points on the potential energy surface should they be of interest for a computational study.

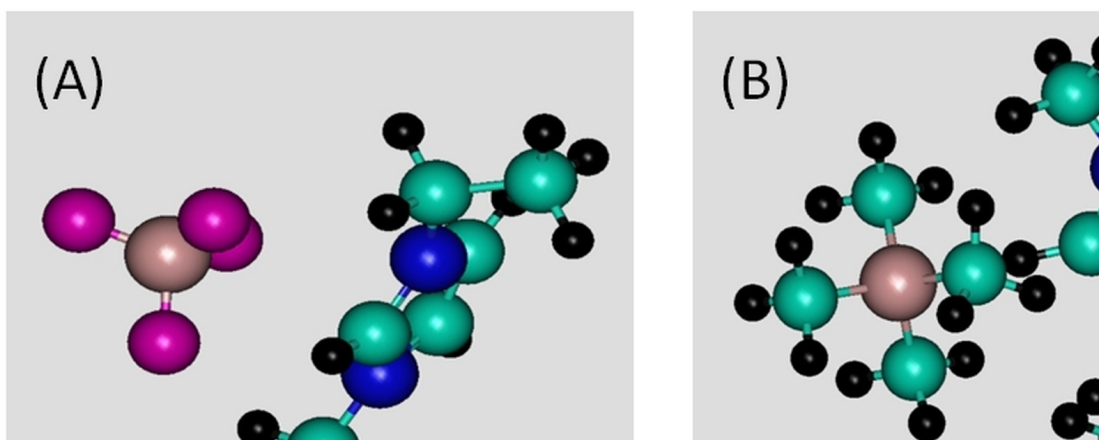

Figure 2. Global minimum configurations of (A) 1-ethyl-3-methylimidazolium tetrafluoroborate and (B) 1-ethyl-3-methylimidazolium tetramethylborate ion pairs. Boron atoms are pale pink; carbon atoms are cyan; nitrogen atoms are blue; fluorine atoms are purple; hydrogen atoms are black.

The stationary points recorded in the [EMIM][Et$_4$B] and [EMIM][Ph$_4$B] systems are given in Figure 3, whereas the selected ionic configurations are exemplified in Figure 4. One notices that larger systems feature larger energetic differences among the local minimum states, while their total number also increases somewhat. Several transition points above +200 kJ mol$^{-1}$ reveal repacking of ions upon searching for alternative stable configurations.

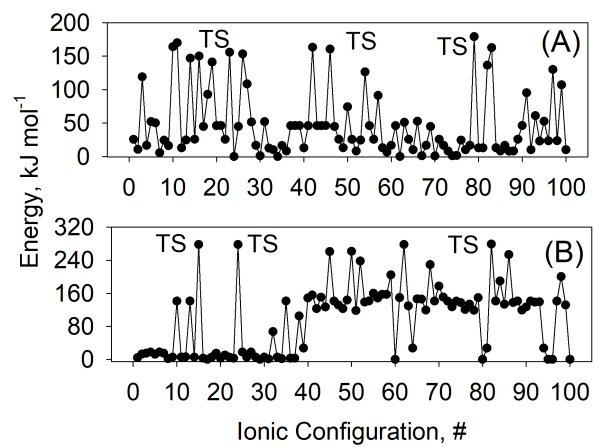

Figure 3. The potential energies of the stationary points found on the potential energy surface of (A) 1-ethyl-3-methylimidazolium tetraethylborate and (B) 1-ethyl-3-methylimidazolium tetraphenylborate ion pairs. The energy of the global potential minimum is set to zero.

The tetraphenylborate anion induces fundamental changes in the cation-anion coordination (Figure 4). Indeed, the most acidic hydrogen atom of the imidazole ring does not play a major role in the interionic attraction based on the global minimum configuration. In the absence of the hydrogen...fluorine H-bond the cation prefers an orientation in which two less polarized hydrogen atoms of the imidazole ring are closer to the anion. Furthermore, the new global minimum conformation fosters maximization of the London attraction between the involved ions.

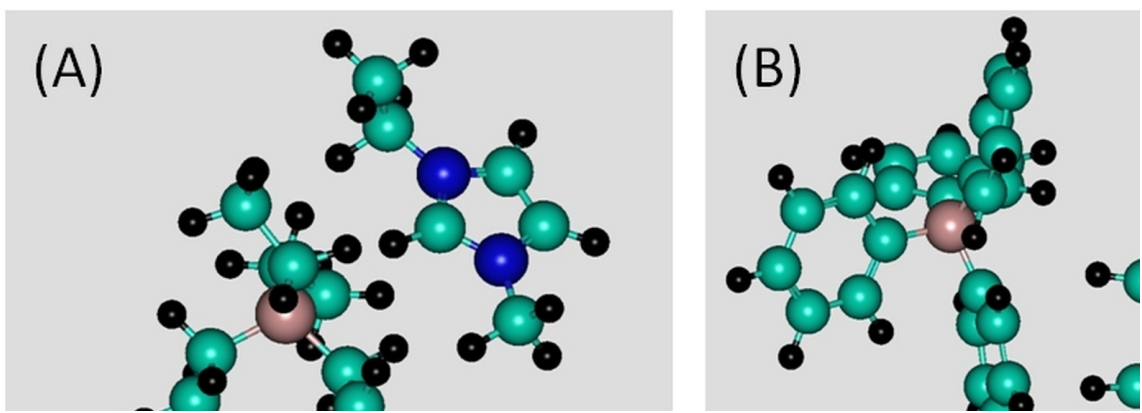

Figure 4. Global minimum configurations of (A) 1-ethyl-3-methylimidazolium tetraethylborate and (B) 1-ethyl-3-methylimidazolium tetraphenylborate ion pairs. Boron atoms are pale pink; carbon atoms are cyan; nitrogen atoms are blue; hydrogen atoms are black.

Table 1 summarizes the most representative distances to describe the cation-anion electrostatic and van der Waals interactions in the most energetically favorable ionic configurations of the considered systems. Whereas the hydrogen bond exists only in the [EMIM][BF$_4$] system, the closest-approach distances remain similar for larger anions. For instance, the smallest distance between the anion and the nitrogen atom of the imidazole ring in [EMIM][B(C$_2$H$_5$)$_4$] amounts to 0.300 nm which is even smaller than the same distance in [EMIM][BF$_4$]. However, the chemical difference is essential. No hydrogen bond exists in [EMIM][B(CH$_3$)$_4$], [EMIM][B(C$_2$H$_5$)$_4$], and [EMIM][B(C$_6$H$_5$)$_4$]. Therefore, the cation-anion attraction forces get must smaller. This is clearly evidenced by the number of stationary points located in the considered systems. The absence of a single preferable coordination pattern gives rise to novel states and more complex molecular dynamics in the condensed phase.

Table 1. Geometrical parameters of the cation-anion electrostatic interaction represented by the smallest non-covalent interatomic distances.

| System | Anion–H (acidic), nm | Anion–H (methyl), nm | Anion–H (ethyl), nm | Anion–N (ring), nm |
|---|---|---|---|---|
| [EMIM][BF$_4$] | 0.191 | 0.226 | 0.222 | 0.304 |
| [EMIM][B(CH$_3$)$_4$] | 0.198 | 0.233 | 0.218 | 0.302 |
| [EMIM][B(C$_2$H$_5$)$_4$] | 0.205 | 0.255 | 0.239 | 0.300 |
| [EMIM][B(Ph)$_4$] | 0.607 | 0.265 | 0.284 | 0.424 |

It is important to note that the cation-anion coordination in the [EMIM][B(Ph)$_4$] system (global minimum configuration) is principally different from all other ion pairs according to the cation's primary interaction site. As a result, the distance between the intrinsically acidic hydrogen atom of the imidazole ring and the closest atom of the [B(Ph)$_4$] anion is 0.607 nm. The two remaining hydrogen atoms of the imidazole ring are located closer to the anion in their equilibrium configuration.

Figure 5 investigates how the size of the anion influences the distance between the boron atom and the acidic hydrogen atom of the cation. This information is essential to describe an effect of the chemically inert hydrocarbon chains that substituted the polarized fluorine atoms. The impact of the methyl and ethyl substituents is negligible as compared to the fluorine atoms, whereas the effect of phenyl is much stronger. The phenyl chains perfectly protect the center of the anion from the polar moieties of the imidazole ring. As a result, an electrostatic component of the cation-anion binding is substantially decreased.

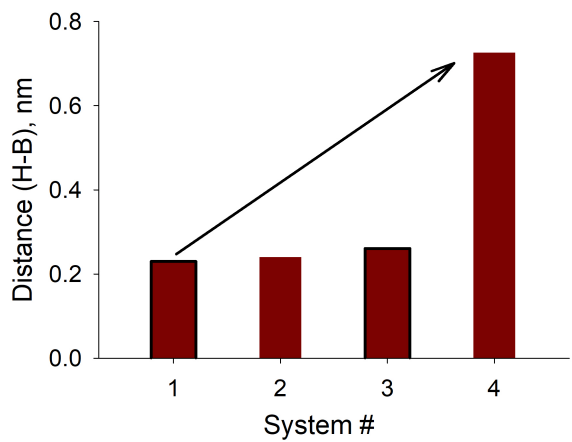

Figure 5. The distance between the center of the anion (boron atom) and the most reactive center of the imidazolium-based cation (acidic carbon atom).

The cation-anion coordination is usually accompanied by a partial electron transfer and electronic polarization of both ions that lead to the redistribution of partial electrostatic charges over the ions. Figures 6 and 7 provide total charges per ion and the boron atom partial charge computed according to the CHELPG (by Breneman and Wiberg)[43] and Charge-Model-5 (by Marenich and coworkers)[44] algorithms as a function of the anion's size.

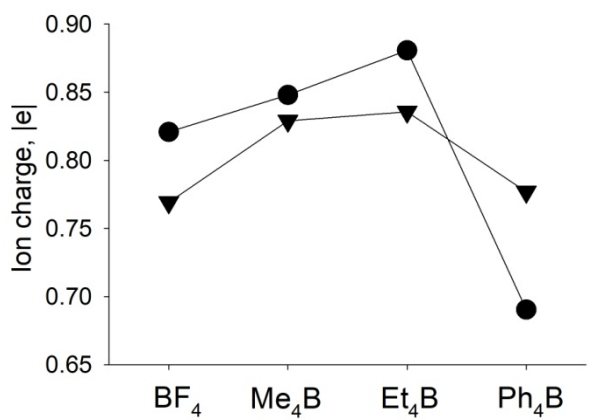

Figure 6. The sum of all partial charges on an ion in the simulated ion pair in the global minimum configuration as a function of the anion employed. The CHELPG charges are circles, whereas the CM5 charges are triangles.

The total cation's charge increases when the fluorine atoms are substituted by the methyl and ethyl groups. Evidently, the substituents blocking the approach of the polar interaction centers to one another decrease the percentage of electron transfer in the respective ion pair systems. The phenyl groups, however, make an exception. The sum of charges per ion decreases significantly. We find this computational observation to be in line with the properties of the pi-electron subsystem of benzene. The higher-energy electrons of the phenyl groups foster valence electron sharing that ultimately results in a smaller total charge on both adjacent ions.

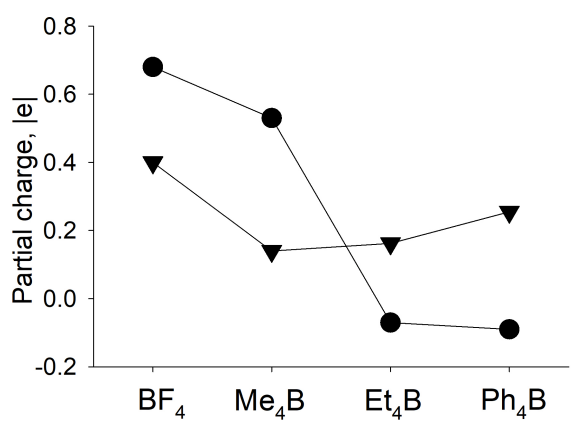

Figure 7. CHELPG (circles) and CM5 (triangles) partial charges on the boron atom in the global minimum configurations of the 1-ethyl-3-methylimidazolium salts depending on the anion employed.

Our above hypothesis gets its confirmation from Figure 7 that summarizes the partial charges on the boron atoms. The boron atom is not sensitive to the chemical nature of the substituents but it is sensitive to the distance of the center of the anion to the imidazole ring. The partial charges that were derived from the electrostatic potential fitting procedure decrease as the size of the substituent group increases. In turn, the CM5 charges are quite insensitive to the substituent. Whereas the partial charge on boron is +0.40 a.u. in the tetrafluoroborate anion, it is +0.26 a.u. in [Ph$_4$B]. This difference is extremely small in comparison with the difference in bulkiness, number of atoms and electrons in the considered anions.

The results of our work include not only global minimum configurations but also a rich set of stationary points. These states also contribute to the macroscopic physical-chemical properties of the studied ionic systems and need to be scanned for the sake of finding kinetically stable and characteristic structural patterns. Figure 8 visualizes some ionic configurations that are local minima. They differ from the respective global minima states by somewhat altered non-covalent geometrical parameters (angles, dihedrals). In most cases, such geometrical alterations can be linked to rotations of certain electrostatically interacting moieties of the cation or the anion or to the inherent conformational flexibility of each ion per se. Note the local minimum configuration in [EMIM][Ph$_4$B] corresponds to the most acidic hydrogen atom of the imidazole ring directly participating in the coordination of the neighboring anion.

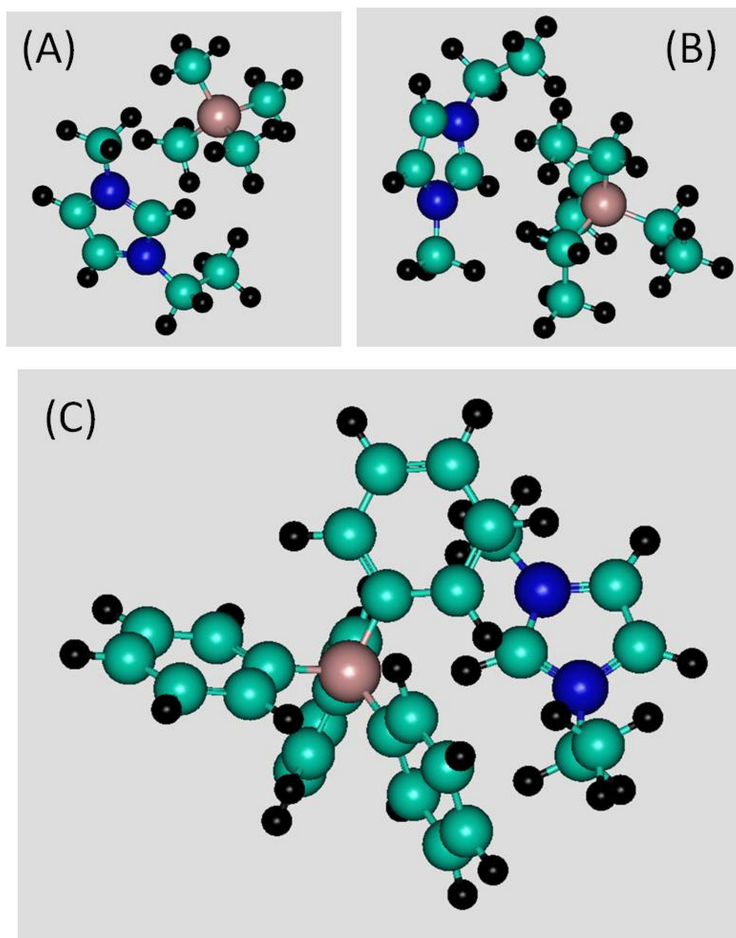

Figure 8. Selected local minima ionic configurations of (A) [EMIM][Me$_4$B], (B) [EMIM][Et$_4$B], and (C) [EMIM][Ph$_4$B]. The relative potential energies of the configurations are +13 kJ mol$^{-1}$, +24 kJ mol$^{-1}$, +18 kJ mol$^{-1}$, respectively. Boron atoms are pale pink; carbon atoms are cyan; nitrogen atoms are blue; hydrogen atoms are black.

The size of the system is an important descriptor of the simulation. The choice of the adequate system's size depends on the investigated phenomenon. Certain physical properties are determined by collective correlation functions and their correct simulation requires a significantly large system to be computationally described. In turn, the local structure properties can, in most cases, be reproduced by a few particles. This is the reason why the classical force field models developed on the basis of very small systems exhibit satisfactory accuracy even when applied to the simulation of collective correlations of the microscopic particles to obtain physical properties, such as thermal conductivity, ionic conductivity, shear viscosity, etc.

In the present work, it is important to identify the role of the system's size and demonstrate that the systems (Table 1) used to study the potential energy surfaces of the ionic systems can be successfully applied. Figure 9 provides the results of the potential energy surface search, whereas Figure 10 exemplifies the global minimum ionic configurations obtained for the twice larger system of [EMIM][Me$_4$B] composed of four ions.

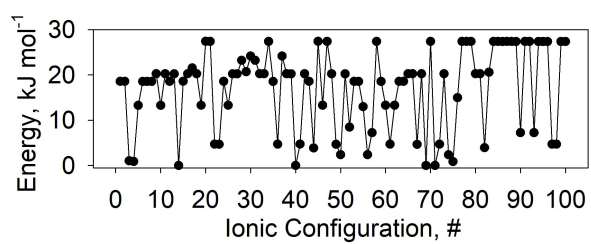

Figure 9. Relative potential energies of local minima in the [EMIM][Me$_4$B] system (two ion pairs). The energy of the global potential minimum has been set to zero.

The number of the available stationary points depends on the number of degrees of freedom in the system. The latter depend on the size of the systems in terms of atoms. A larger system contains a proportionally larger number of local minima and stationary points. Therefore, the number of stable configurations is an extensive property and, hence, larger systems require more extensive sampling procedures. The most frequently observed structural patterns in the systems of different size but the same chemical composition can be smoothly compared to one another.

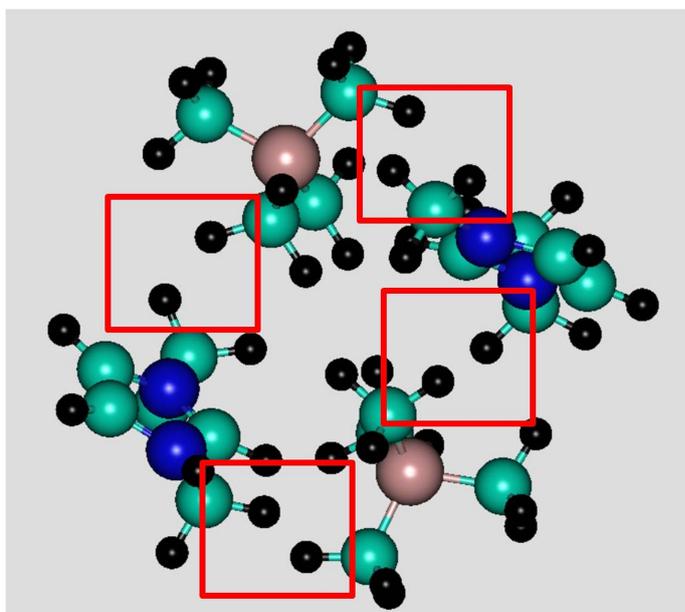

Figure 10. The global minimum ionic configuration of [EMIM][Me$_4$B] comprising two cations and two anions. Red squares underline the closest-approach distances between the cation and the anion. Boron atoms are pale pink; carbon atoms are cyan; nitrogen atoms are blue; hydrogen atoms are black.

Figure 10 proves that the cation-anion interaction pattern (coordination of the non-polar hydrogen atoms) in [EMIM][Me$_4$B] is reproduced both in the smaller and larger systems. The role of the most acidic hydrogen atom of the imidazole ring clearly decreases that is confirmed by a higher conformational flexibility in the case of larger anions and should lead to the increase of the ionic transport properties. Thus, the initially selected size of the systems allows for obtaining a trustworthy description of the local structural interionic patterns.

In the case of the discussed larger system, we used a softer kinetic energy perturbation. Unlike for other systems, the perturbation was equivalent to 500 K. The procedure was modified because we were not interested in sampling transition states. Instead, we rather aimed to get local minimum states to compare them with the analogous states in the smaller system. As a result, Figure 9 does not contain any transition points. It should be noted that the magnitude of the perturbation in the applied sampling method acts as a limiting factor for the investigation of the potential energy surface. For instance, a rather modest periodic perturbation used in this work does not allow for any chemical reactions to be initiated. The applied kinetic energy perturbation, therefore, must be chosen responsibly and with a phenomenon of interest in mind.

**Conclusions**

The physical-chemical properties of RTILs represent a many-dimensional function wherein the constituent ions are the cornerstone parameters. The cation-anion interaction heavily depends on the specific coordination sites available in the structures of both involved ions. In this work, we showed that the adjustment of interionic attraction is possible via a straightforward chemical modification of the anion's structure. The modified tetraalkylborate anions foster redistribution of the electron density over all ions, change the major structural pattern and, most importantly, eliminate a strong cation-anion electrostatic attraction that leads to a hydrogen bond formation in 1-ethyl-3-methyl-imidazolium tetrafluoroborate.

The systematic investigation of stationary points is equivalent to the most comprehensive sampling of the phase space in molecular dynamics simulations. Furthermore, an ability to observe ionic configurations that are located far from the global minimum state assures that we considered all relevant points in the simulated systems. The applied method forms a potentially fruitful basis to succeed in the target molecular design and the development of novel task-specific ionic liquids.

In the future, the experimental measurement must be performed to identify an extent to which the melting temperature in the imidazolium-based RTILs decreases thanks to the removal of the strongest interionic attraction pattern that involves an intrinsically acidic hydrogen atom of the positively charged imidazole ring.

**Acknowledgments**

All reported numerical simulations have been conducted at the P.E.S. computational facility.

**Conflict of interest**

The author does not have financial or personal relationships that might inappropriately influence (bias) their actions: dual commitments, competing interests, or competing loyalties.

**Authors for correspondence**

All correspondence regarding the content of this paper shall be directed through electronic mail to the author: vvchaban@gmail.com (Professor Dr. Vitaly V. Chaban).